\documentclass[12pt,twoside]{article}
\usepackage{graphicx}
\usepackage{cmp2e}
\title[Critical slowing down in random anisotropy magnets]%
{Critical slowing down in random anisotropy magnets%
}

\author[M. Dudka,
        R. Folk, Yu. Holovatch,
        G. Moser]{M. Dudka\refaddr{label1,label2},
        R. Folk\refaddr{label2}, Yu. Holovatch\refaddr{label1,label2,label3},
        G. Moser\refaddr{label4} }
\addresses{
 \addr{label1} Institute for Condensed Matter Physics, National
Academy of Sciences of Ukraine, UA--79011 Lviv, Ukraine
 \addr{label2} Institute f\"ur Theoretische Physik,
Johannes Kepler Universit\"at Linz, A-4040, Linz, Austria
 \addr{label3} Ivan Franko National University of Lviv,
UA-79005 Lviv, Ukraine
 \addr{label4} Institute f\"ur
Physik und Biophysik, Univerit\"at Salzburg, A-5020 Salzburg,
Austria}

\begin{document}

\maketitle

\begin{abstract}
We study the purely relaxational critical dynamics with
non-conserved order parameter  (model A critical dynamics) for
three-dimensional magnets with disorder in a form of the random
anisotropy axis. For the random axis anisotropic distribution, the
static asymptotic critical behaviour coincides with that of random
site  Ising systems. Therefore the asymptotic critical dynamics is
governed by the dynamical exponent of the random Ising model.
However, the disorder influences considerably the  dynamical
behaviour in the non-asymptotic regime. We perform a
field-theoretical renormalization group analysis within the
minimal subtraction scheme in two-loop approximation to
investigate asymptotic and effective critical dynamics of random
anisotropy systems. The results demonstrate the non-monotonic
behaviour of the dynamical effective critical exponent $z_{\rm
eff}$.
\pacs 05.50.+q; 05.70.Jk; 61.43.-j; 64.60.Ak;  64.60.Ht;
\end{abstract}

\section{Introduction} \label{I}
The concept of scaling plays a central role in  modern theory of
critical phenomena   \cite{books}. Introducing a set of appropriate
scaling variables a large amount of experimental and numerical
data can be described by a few scaling functions   \cite{scaling}.
The most prominent effect in
dynamical critical phenomena is {\em critical slowing
down} which consists in an increase of the relaxation time
approaching the critical point. This is induced by the divergence
of the correlation length $\xi$ at the critical point and causes
also relaxation time $\tau$ to diverge with the  dynamical critical
exponent $z$:
\begin{equation}\label{1}
\tau\sim\xi^z.
\end{equation}
Renormalization group  (RG) theory, with its concepts of
invariance of the system at the critical point against changes in
length scale gives a basis to universality in connection with
fixed points and static   \cite{RGbooks} and dynamic scaling
  \cite{Halperin77}.

This holds however only in the asymptotic region in the vicinity
of the critical point. Further away scaling breaks down and the
description of critical phenomena becomes more complicated and
involves non-universal characteristics both in statics and in
dynamics   \cite{books,Yukhnovskii}.

Another important point for observing the behaviour in certain
universality classes is the homogeneity of the system under
consideration. Therefore the effect of impurities on critical
behaviour is of considerable interest.  However it turned out the
disordered systems may show also scaling within a certain
universality class. This might be the universality class of a pure
system or a  new one
  \cite{Harris74,Boyanovsky82,Weinrib83,Luck93}. Moreover the
changes introduced by the disorder depend on the type of this
disorder; namely is it introduced by dilution (random site
  \cite{diluted} or random bond   \cite{Berche04} systems), or as a
random field   \cite{Belanger91}, random connectivity
  \cite{Luck93,Janke04} or as an anisotropy   \cite{ramreviews}. The
defects may be correlated   \cite{Boyanovsky82,Weinrib83} or not. It
may even happen that the second order transition of the pure
system is destroyed   \cite{Imry75,Pelcovits78}.

It further turned out that considering a specific system with
defects the behaviour near the critical point seems to be
non-universal. Knowing that the non-asymptotic behaviour is
non-universal it became necessary to study the non-asymptotic
behaviour of such systems in more detail. Indeed in many systems
(e.g with site disorder) effective critical behaviour can explain
the experimental situation   \cite{Berche04,effective}.

RG investigations of dynamic critical behaviour is in many cases
technically  much more involved in comparison to
statics. As a consequence results for dynamics are known
and in much lower approximations (in most cases only up to two loop order).
On the other hand for the dynamics of magnetic systems with mode coupling terms the dynamical
critical exponent is known  exactly or contains only a static exponent.

In this paper, we will present an analysis of the dynamical
critical behaviour of random anisotropy magnets which constitute a
large class of disordered systems   \mbox{\cite{ramreviews}}. In order to
give some examples, the majority of the amorphous rare-earth
containing alloys is recognized as random anisotropy magnets
  \cite{ramreviews}, certain crystalline compounds with an
rear-earth component belong to this class too   \cite{Moral}. Random
anisotropy characterizes also molecular based magnets
  \cite{molmag}, nanocrystalline materials   \cite{nano}, as well as
granular systems   \cite{granular}. Moreover, the analysis involving
random anisotropy found its application also in  the
interpretation of the phase transition in liquid crystals in
porous media   \cite{liqcryst}.

The model currently used for the description of random anisotropy
systems was introduced in the early 70-ies by Harris, Plischke,
and Zuckermann   \cite{Harris73}.  It describes $m$-component spins
located on the sites of a $d$-dimensional lattice, each spin
subjected to a local anisotropy of random orientation. The
Hamiltonian of the random anisotropy model (RAM) reads:
\begin{equation}
{\mathcal H} =  - \sum_{{\bf R},{\bf R'}} J_{{\bf R},{\bf R'}}
\vec{S}_{\bf R} \vec{S}_{\bf R'} -D_0\sum_{{\bf R}} (\hat {x}_{\bf
R}\vec{S}_{\bf R})^{2}.
 \label{origham}
\end{equation}
Here,  $\vec{S}_{\bf R}=(S^1_{\bf R},\dots,S^m_{\bf R})$, vectors
${\bf R}$ span sites of a $d$--dimensional cubic lattice, $D_0>0$
is an anisotropy constant, $\hat {x}$ is a random unit vector
specifying direction of the local anisotropy axis. The interaction
$J_{{\bf R},{\bf R'}}$ is assumed to be ferromagnetic. Note, that
for the Ising-like magnets, $m=1$, the last term in
(\ref{origham}) is just a constant, therefore the random
anisotropy is present for $m>1$ only.

Below, we will consider quenched disorder, when the vectors $\hat
{x}_{\bf R}$ in (\ref{origham}) are randomly distributed with a
distribution function $p(\hat x)$ and fixed in a certain
configuration. It is well established by now, that the anisotropy
axis distribution plays a crucial role for the origin of the
low-temperature phase in the RAM. In particular, when the random
vectors $\hat {x}_{\bf R}$ point with equal probability towards
any direction, such a distribution may be called an {\em isotropic} one,
the ferromagnetic ordering is impossible   \cite{isotropic} for
spatial dimension $d\leq4$. Whereas it may occur for an
anisotropic distribution. In statics, this situation was
corroborated by the RG studies of RAM
  \cite{Aharony75,Mukamel82,Korzhenevskii88,RAM,Calabrese04}
restricting $\hat x$ to point along one of the $2m$ directions of
the axes $\hat k_i$ of a hypercubic lattice  ({\em cubic distribution}):
\begin{equation}\label{cubdist}
p(\hat x) = \frac{1}{2m}\sum_{i=1}^m\left[\delta^{(m)}(\hat x-\hat
k_i)+\delta^{(m)}(\hat x+\hat k_i)\right],
\end{equation}
with  Kronecker deltas  $\delta(y)$. In this case, the second
order phase transition into the magnetically ordered low-temperature
phase occurs and in asymptotics it is characterized by the
critical exponents of the random-site Ising model, the fact
suggested already in Ref.   \cite{Mukamel82} and confirmed later in
Refs.   \cite{Korzhenevskii88,RAM,Calabrese04}.

Taken that the studies of static criticality of random anisotropy
magnets are far from being as intensive as those of the diluted
magnets   \cite{diluted}, even  less is known about their dynamic
critical behaviour. The  dynamical models for systems with
{\em isotropic} distribution of local anisotropy axis were considered in
Refs.   \cite{Ma78,DeDominicis78,Khurana82}, the first order RG
calculations were performed in Ref.  \cite{Krey77}. Whereas the
problem of dynamic critical behaviour of a RAM with an anisotropic
random axis distribution as to our knowledge up to now remained
untouched.

In this paper, we consider a purely relaxational dynamics of a
three-dimensional ($d=3$) RAM with non-conserved order parameter
(model A in classification of Ref.  \cite{Halperin77}) and a {\em cubic}
random axis distribution (\ref{cubdist}). As far as the static
critical behaviour of such magnets (note, with $m>1$) belongs to
the universality class of a random-site Ising model  \cite{Mukamel82,Korzhenevskii88,RAM,Calabrese04}, for which the
heat capacity does not diverge  \cite{diluted}. Therefore, the
critical dynamics of such a model is governed in asymptotics by
the model A random-site Ising magnet dynamical critical exponent
for any order parameter components number $m$. However in the
non-asymptotic region, the model possesses a rich effective
critical behaviour, as will be shown by our subsequent analysis.
Taken that it is this  effective behaviour which is observed both
in experiments and in the MC simulations, it is important to have
a RG prediction for typical scenarios of the  approach to
criticality in  random anisotropy magnets.

The rest of the paper is organized as following: in the next
section \ref{II} we present the Langevin equations governing model
A dynamics and describe the renormalization procedure, in section
\ref{III} we give results of our calculations obtained in two-loop
approximation and display possible scenarios of effective
critical behaviour. Conclusions and outlook are given in section
\ref{IV}.

\section{Model equations and renormalization} \label{II}

We consider the dynamics for model (\ref{origham}) with random axis
distribution (\ref{cubdist}) to be relaxational with non-conserved
$m$-component order parameter $\vec{\varphi}_0 \equiv
\vec{\varphi}_0(R)$. In this case the relaxation of the order
parameter is described by the Langevin equation:
\begin{eqnarray}\label{eq_mov2}
\frac{\partial {\varphi}_{i,0}}{\partial
t}&=&-\mathring{\Gamma}\frac{\partial {\mathcal H}}{\partial
{\varphi}_{i,0}}+{\theta}_{{\varphi}_i}\qquad i=1\ldots m,
\end{eqnarray}
with the Onsager coefficient $\mathring{\Gamma}$,  stochastic
forces  ${\theta}_{\varphi_i}$  obeying the Einstein relations:
\begin{eqnarray}\label{11}
<{\theta}_{\varphi_i}(R,t){\theta}_{\varphi_j}(R',t')>&
=&2\mathring{\Gamma}\delta(R-R')\delta(t-t')\delta_{ij},
\end{eqnarray}
and the disorder-dependent equilibrium effective Hamiltonian
$\mathcal H$:
\begin{eqnarray}
\label{effram}\!\!\!\!\!{\cal H}&{=}\!\!& \int\! d^d R
\left\{{1\over 2} \left[|\nabla \vec{\varphi}_0|^2{+}\mathring{r}
|\vec{\varphi}_0|^2\right]\!{+} \frac{v_0}{4!}
|\vec{\varphi}_0|^4{-}\! D\!\left( \hat x
\vec{\varphi}_0\right)^2\right\}.
\end{eqnarray}
In (\ref{effram}), the field $\vec\varphi_0$ is an $m$-component
vector, $D$ is proportional to the anisotropy constant of the spin
Hamiltonian (\ref{origham}) with $D_0$; $\mathring{r}$ and $v_0$ are
defined by $D_0$ and the fourth order coupling of the $m$-vector magnet
(see Ref.  \cite{RAM} for details).

We treat the critical dynamics of the disordered model within the
field theoretical RG method based on the Bausch-Janssen-Wagner
formulation  \cite{Bausch76}, where the appropriate Lagrangians are
studied. For the model equations (\ref{eq_mov2})-(\ref{11}) the Lagrangian
reads:
\begin{equation}
{\cal L}=\int d^d R
dt\sum_i\tilde\varphi_{i,0}\left[\frac{\partial
\varphi_{i,0}}{\partial t}+\mathring{\Gamma}
 \frac{\delta {\cal H}}{\delta
 \varphi_{i,0}}-\mathring{\Gamma}\tilde\varphi_{i,0}\right],
\end{equation}
with a new auxiliary response $m$-component field
${\tilde{\vec\varphi}}_0$. Here and below sums over field
components span values from 1 to $m$.

Studying critical properties of disordered systems one should
average over the random degrees of freedom. In order to treat
quenched disorder, in statics often the replica trick is 
used  \cite{diluted}. However, it was established in  \cite{DeDominicis78}
that in dynamics it is not necessary to make use of the replica
trick: it is sufficient to the  average over the random variables
$\hat x$ with their distribution (\ref{cubdist}). Then the Lagrangian
for the model reads:
\begin{eqnarray}\label{rLagrangian}
{\mathcal L}\!&=& \! \Bigg\{\int d^d R dt
\sum_i{\tilde{\varphi}_{i,0}}\Bigg[
 \left({\frac{\partial}{\partial t}}{+}\mathring{\Gamma}(r_0{-}
\nabla^2)\right)\varphi_{i,0}{-}\mathring{\Gamma}{\tilde{\varphi}_{i,0}}
 {+}\frac{\mathring{\Gamma}{v_0}}{3!}{\varphi_{i,0}}\sum_j{\varphi}_{j,0}{\varphi_{j,0}}
 {+}\frac{\mathring{\Gamma}{y_0}}{3!} {\varphi^3_{i}}\Bigg]{+}\nonumber\\ &&\int
dt'\sum_{i}\tilde{\varphi}_{i,0}(t) {\varphi_{i,0}}(t)\Bigg[
\frac{\mathring{\Gamma}^2{u_0}}{3!}\sum_{j}\tilde{\varphi}_{j,0}(t')
{\varphi_{j,0}}(t'){+}\frac{\mathring{\Gamma}^2{w_0}}{3!}\tilde{\varphi}_{i,0}(t')
{\varphi_{i,0}}(t')\Bigg]\Bigg\}.
\end{eqnarray}
Here, $r_0$ is proportional to the temperature distance to the  mean field
critical point and the bare couplings are $u_0>0$, $v_0>0$, $w_0<0$.
Moreover, $u_0$ and $w_0$ are connected to the moments of the
distribution (\ref{cubdist}) in such a way that $w_0/u_0=-m$. The
term with $y_0$ does not appear after averaging over disorder,
however it should be added since it will be generated within the
renormalization procedure. The set of static couplings $\{u_0,
v_0, w_0,y_0 \}$ will be denoted below by $\{u_{0,i} \}, \,
i=1,..,4$.

\vspace*{2ex}
\begin{figure}[htbp]
\begin{picture}(200,100)
\put(20,90){\includegraphics[width=0.1\textwidth]{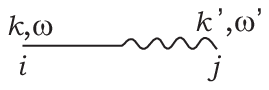}}\put(120,100)
{$G(k,\omega)\delta(k+k')\delta(\omega+\omega')\delta_{i,j}$}
\put(20,70){\includegraphics[width=0.1\textwidth]{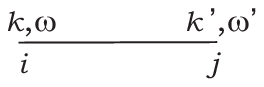}}\put(120,80)
{$C(k,\omega)\delta(k+k')\delta(\omega+\omega')\delta_{i,j}$}
\put(20,35){\includegraphics[width=0.1\textwidth]{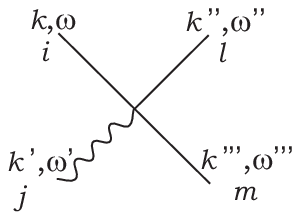}}\put(120,50)
{$\Gamma
A\delta(k+k'+k''+k''')\delta(\omega+\omega'+\omega''+\omega''') $}
\put(20,0){\includegraphics[width=0.13\textwidth]{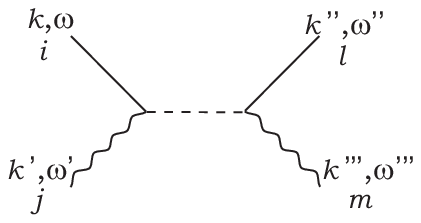}}\put(120,15)
{$\Gamma^2 B
\delta(k+k')\delta(k''+k''')\delta(\omega+\omega')\delta(\omega''+\omega''')
$}
\end{picture}
\caption{\label{elem} Elements for construction of Feynman
diagrams. The response function and the correlation function read:
$G(k,\omega) = 1/(-i\omega+\mathring{\Gamma}(r+k^2))$,
$C(k,\omega) =
2\mathring{\Gamma}/|-i\omega+\mathring{\Gamma}(r+k^2)|^2$. $A$
stands for ${v_0}/{3!}\,
{(\delta_{i,j}\delta_{l,m}+\delta_{i,l}\delta_{j,m}+\delta_{i,m}\delta_{j,l})}/{3}$
or for ${y_0}/{3!}\,\delta_{i,j}\delta_{j,l}\delta_{l,m}$ while
$B$ equals to ${u_0}/{3!}\,\delta_{i,j}\delta_{l,m}$ or to
${w_0}/{3!}\,\delta_{i,j}\delta_{j,l}\delta_{l,m}$. }
\end{figure}

\begin{figure}[htbp]
\begin{picture}(400,130)
\put(70,100){\includegraphics[width=0.17\textwidth]{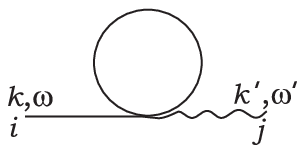}}\put(230,100)
{\includegraphics[width=0.18\textwidth]{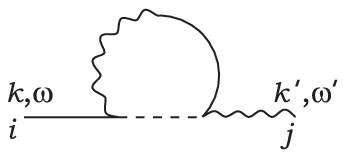}}
\put(0,35){\includegraphics[width=0.18\textwidth]{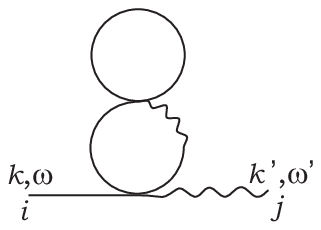}}\put(80,35)
{\includegraphics[width=0.18\textwidth]{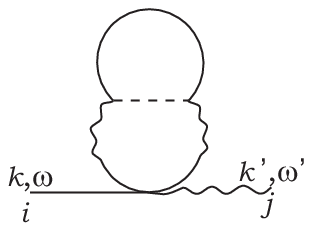}}
\put(160,35){\includegraphics[width=0.18\textwidth]{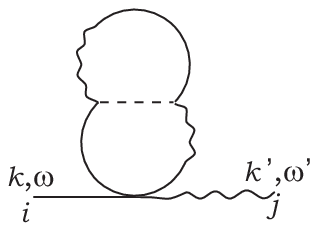}}\put(240,35)
{\includegraphics[width=0.18\textwidth]{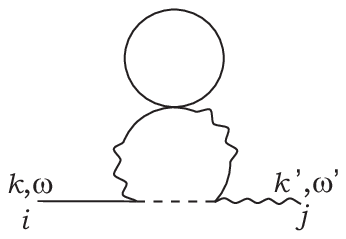}}
\put(320,35){\includegraphics[width=0.18\textwidth]{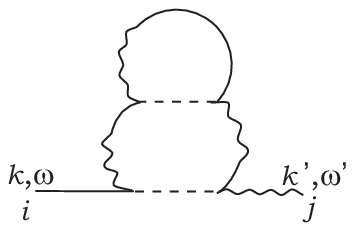}}\put(0,0)
{\includegraphics[width=0.18\textwidth]{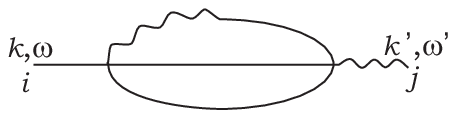}}
\put(100,0){\includegraphics[width=0.18\textwidth]{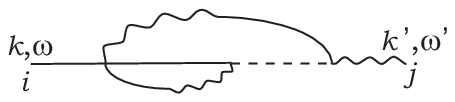}}\put(200,0)
{\includegraphics[width=0.18\textwidth]{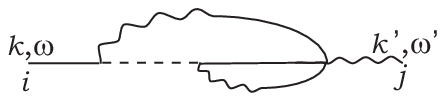}}
\put(300,0){\includegraphics[width=0.18\textwidth]{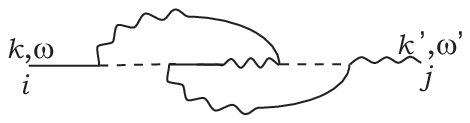}}
\end{picture}
\caption{\label{oldiag} Diagrams of the function
${{\mathring{\Gamma}_{\tilde{\varphi}\varphi}}}^{i,j}({r}_0,
\{ u_{i,0}\},\mathring{\Gamma},k,\omega)$ up to two-loop order. First two terms represent
one-loop contribution, while the rest of diagrams is of two-loop
order.}
\end{figure}

With the Lagrangian (\ref{rLagrangian}) depending on the bare
quantities (denoted by the sub- and superscripts 'o') we analyze
within  field theory  \cite{RGbooks} the dynamical vertex
functions. As far as the static RG functions have been obtained
before  \cite{Dudka05,RAM}, we need only to calculate  the
two-point dynamical vertex function
${{\mathring{\Gamma}_{\tilde{\varphi}\varphi}}}^{i,j}({r}_0,
\{ u_{i,0}\},\mathring{\Gamma},k,\omega)=
{\mathring{\Gamma}_{\tilde{\varphi}\varphi}}({r}_0, \{
u_{i,0}\},\mathring{\Gamma},k,\omega)\delta_{i,j}$. The
calculations are performed using Feynman diagrams, elements for
them are given in the Fig. \ref{elem} whereas the one- and
two-loop contributions to
${{\mathring{\Gamma}_{\tilde{\varphi}\varphi}}}^{i,j}$ are
depicted in Fig. \ref{oldiag}.

We perform renormalization of
${\mathring{\Gamma}_{\tilde{\varphi}\varphi}}$
 within the minimal subtraction scheme  \mbox{\cite{RGbooks}}. In this
 scheme, in order to define renormalized static ($\varphi$, $r$,
$\{u_i\}$) and dynamic ($\tilde\varphi$, $\Gamma$) fields and
couplings, the renormalization factors $Z_{\varphi}$, $Z_r$,
$Z_{u_i}$ and $Z_{\tilde\varphi}$, $Z_{\Gamma}$ are introduced by:
\begin{equation} \label{zfactors}
\varphi=Z^{-1/2}_{\varphi}\varphi_0, \,
\tilde\varphi=Z^{-1/2}_{\tilde\varphi}{\tilde\varphi}_0, \,
r=Z_r^{-1}r, \, u_i=\mu^{-\varepsilon}Z_{u_i}^{-1}Z^2_{\varphi}A_d
u_{0,i}, \, \Gamma=Z_{\Gamma}\mathring{\Gamma}.
\end{equation}
Here, $\mu$ is the external momenta scale, $\varepsilon=4-d$, and
$A_d$ is a geometrical factor.

The critical behaviour of the system is described by the following
RG functions:
\begin{equation} \label{rgfunctions}
\beta_{u_i}(\{u_i\})=\left.\mu\frac{\partial u}{\partial
\mu}\right|_0, \hspace{2em}
\zeta_{r}(\{u_i\})=\left.-\mu\frac{\partial \ln Z_r}{\partial
\mu}\right|_0, \hspace{2em}
\zeta_{\Gamma}(\{u_i\})=\left.\mu\frac{\partial \ln
Z_{\Gamma}}{\partial \mu}\right|_0,
\end{equation}
where the symbol $\left.\frac{}{}\right|_0$ means differentiation at
fixed bare parameters. The $\beta$-functions determine the RG flow
of couplings under renormalization:
\begin{equation}\label{flow}
\ell\frac{d u_i}{d \ell}=\beta_{u_i}(\{u_i\}),\hspace*{3em} i=1,..
,4,
\end{equation}
and the flow parameter $\ell$ is related to the distance from the
critical point. Subsequently, an information about the critical
behaviour of a system can be obtained from analysis of the fixed
points (FPs) of the  flow equations (\ref{flow}). A FP $\{u_i^*\}$
is defined as simultaneous zero of all $\beta$-functions:
\begin{equation}\label{FP}
\beta_{u_i}(\{u^*_i\})=0, \hspace*{3em} i=1,.. ,4.
\end{equation}
The stable and from initial  conditions accessible FP corresponds
to the critical point of the system. A FP is stable if all
eigenvalues $\omega_i$ of the stability matrix
$B_{i,j}=\frac{\partial\beta_{u_i}}{\partial {u_j}}$ calculated at
this FP have  positive real parts.

The FP values of the RG $\zeta$-functions (\ref{rgfunctions})
determine the asymptotic values of critical exponents. In
particular, the dynamical asymptotic critical exponent $z$
(\ref{1}) is given at the  stable and accessible FP by:
\begin{equation}\label{asym}
z=2+\zeta_{\Gamma}(u^*,v^*,w^*,y^*).
\end{equation}
While the effective dynamical exponent $z_{\rm eff}$ is calculated
in the non-asymptotical region, where the renormalized couplings did
not reach their FP values and it is defined by the
solutions of the flow equations  (\ref{flow}):
\begin{equation}\label{effec}
z_{\rm eff}=2+\zeta_{\Gamma}(u(\ell),v(\ell),w(\ell),y(\ell)) \, .
\end{equation}
We neglect contributions to $z_{\rm eff}$ coming from the
amplitude function because they are considered to be small.

\section{Results}\label{III}

The static RG functions within the minimal subtraction scheme are
 known in two-loop  \cite{Dudka05} approximation. Within the
massive renormalization
they have been calculated already in five-loop  \cite{Calabrese04} 
approximation. Calculating the dynamical
function $\zeta_{\Gamma}$ within two-loop order we use the static
RG functions of the same order \mbox{\cite{Dudka05}}. Since series for
these static functions are known to be asymptotic at best we use
Pad\'e-Borel resummation scheme  \cite{Baker78} in details
described in Ref.  \cite{Dudka05}.  The FPs values and solutions of flow
equations  \cite{Dudka05} obtained on the  basis of these functions we use in
present study adding to them  the new two-loop expression for
the function $\zeta_{\Gamma}$. The last is derived from the vertex
function ${{\mathring{\Gamma}_{\tilde{\varphi}\varphi}}}^{i,j}$
discussed in Section \ref{II} and reads:
\begin{eqnarray}
\zeta_{\Gamma
}&=&-\frac{(u+w)}{3}+\frac{(6\ln(4/3)-1)}{24}(y^2+\frac{2}{3}vy+\frac{m+2}{3}v^2)+
\nonumber\\&& \label{main}
\frac{1}{36}\left((m+2)uv+5u^2+5w^2+10uw+3yw+3vw+3uy\right).
 \end{eqnarray}

Solving  the FP equations (\ref{FP}) for the static
$\beta$-functions  at fixed space dimension $d=3$  \cite{Schloms}
results in 16 FPs  \cite{RAM,Dudka05}. The region of physical
importance $u>0$, $v>0$, $w<0$ includes 10 FPs. Below we list the most
interesting FPs from
 them
together with the asymptotic value for the $z$ exponent (for the
numerical values of the FPs coordinates obtained in two-loop
approximation with the help of Pad\'e-Borel resummation see  \cite{Dudka05},
 the value of the $z$ exponent, however, is
calculated by a direct substitution of FP coordinates into Eq.
(\ref{asym})):
\begin{itemize}
\item Gaussian FP I: $u^*=v^*=w^*=y^*=0$; $z(\forall
m)=2$;
\item pure FP II: $v^*\neq 0$,  $u^*=w^*=y^*=0$; $z(m=2)=2.053$, $z(m=3)=2.051$;
\item polymer FP III:  $u^*\neq 0$, $v^*=w^*=y^*=0$; $z(\forall
m)=1.815$;
\item Ising FP V: $y^*\neq 0$, $u^*=v^*=w^*=0$; $z(\forall
m)=2.052$;
\item cubic FP VIII:  $u^*{\neq} 0$, $y^*{\neq} 0$, $u^*{=}w^*{=}0$; $z(m{=}2){=}2.157$,
$z(m{=}3){=}2.042$;
\item Ising FP  X: $u^*\neq 0$, $y^*=-w^*$, $v^*=0$; $z(\forall
m)=2.052$;
\item random Ising FP XV:  $w^*\neq 0$, $y^*=-w^*$,   $u^*=v^*=0$; $z(\forall
m)=2.139$.
\end{itemize}
Here, we keep the FP numbering of 
Refs.  \cite{Dudka05,Aharony75,Mukamel82,RAM,Calabrese04}. From the above
list, only the random Ising (XV) and polymer (III) FPs are stable.
However the polymer FP is not accessible from the physical initial
conditions. This leads to the 
conclusion  \cite{Mukamel82,Korzhenevskii88,RAM,Calabrese04} that the random
Ising FP XV governs the critical behaviour. Therefore, the
$m$-vector magnets with {\em cubic} random axis distribution belong to
the universality class of the random-site Ising  magnets. The
non-asymptotic critical behaviour of the RAM
 differs essentially from that of the random-site model
as was demonstrated in statics in Ref.  \cite{Dudka05}. The same
concerns the non-asymptotic dynamical critical behaviour: the
critical slowing down in RAM is governed by $z_{\rm eff}$ exponent
as explained below.

\begin{figure}[htbp]
{\includegraphics[width=0.6\textwidth]{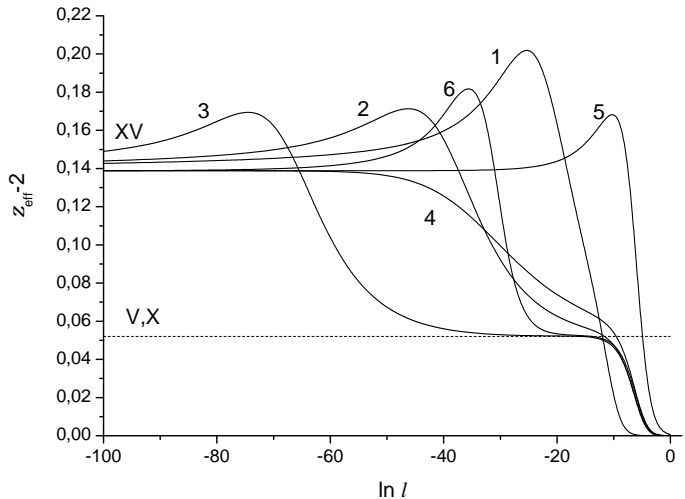}}
\caption{\label{zeff2} Effective critical exponent $z_{\rm eff}$
as a function of the logarithm of the flow parameter for order
parameter dimension $m=2$. Dashed line indicate the value of $z$
at the FPs V, X. See text for details.}
\end{figure}

\begin{figure}[htbp]
{\includegraphics[width=0.6\textwidth]{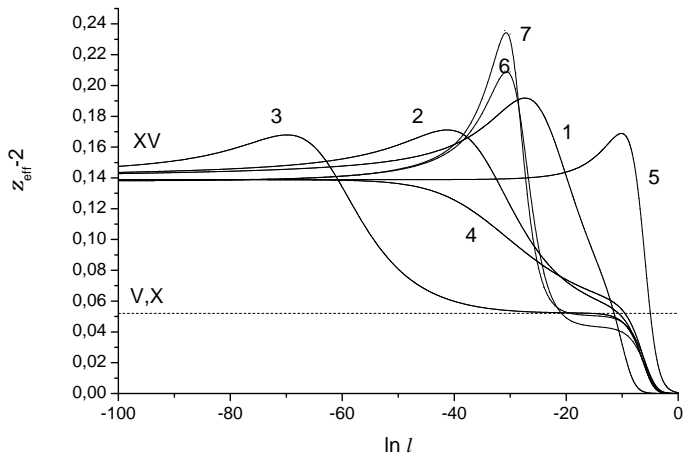}}
\caption{\label{zeff} Effective critical exponent $z_{\rm eff}$ as
a function of the logarithm of the flow parameter for order
parameter dimension $m=3$. Dashed line indicate the value of $z$
at the  FPs V, X. See text for details.}
\end{figure}

The crossovers between different FPs lead to a rich pictures of
possible RG flows  \cite{Dudka05}. Many flows are influenced by the
Ising FPs V and X. Introducing into (\ref{effec}) several typical
RG flows which start from the physical region of initial couplings
one obtains different regimes for approach of the effective
dynamical exponent $z_{\rm eff}$ to asymptotics. The dependence on
the flow parameter $\ell$ of $z_{\rm eff}$ for  easy-plane ($m=2$)
and Heisenberg ($m=3$) magnets are shown in the Figs.~\ref{zeff2}
and~\ref{zeff} correspondingly. Flow 3 was chosen to be affected
by both Ising FPs V and X, therefore both curves 3 of
Figs.~\ref{zeff2} and~\ref{zeff} demonstrate that a large region
for $z_{\rm eff}$ might exist with dynamical exponent values of
the pure one-component (Ising) model A. Curves 6 correspond to
flows which comes near the pure FP II and curve 7 to  flows which
come near the cubic FP VIII.

Although the asymptotic exponents of the random anisotropy magnets
considered here are the same as those of the random-site (diluted)
Ising magnets, the approach to the asymptotical region essentially
differs from the diluted magnets. It is defined by smallest static
stability exponent $\omega=-0.0036$  \cite{Dudka05} which is equal
in absolute value to the ratio of heat capacity critical exponent
$\alpha_r$  and correlations length critical exponent $\nu_r$ of
random-site Ising model  \cite{Calabrese04}. As a consequence the
Wegner correction to scaling is $\Delta=\omega\nu_r=-\alpha_r$.
The high-loop estimate gives $\Delta\approx 
0.049\pm0.009$  \cite{Calabrese04}. Such a small value of $\Delta$ means that the
approach to the asymptotic values is very slow. Therefore
practically only the non-asymptotic critical behaviour governed by
effective critical exponents will be observed experimentally or in
the numerical simulations.

As it is seen from Figs. \ref{zeff2}, \ref{zeff}, another
particular feature of $z_{\rm eff}$  seems  to be that it reaches its
asymptotic value $z$ always from the region $z_{\rm eff}>z$.
 Therefore, in an experimental situation, a
decrease of $z_{\rm eff}$ may serve as an evidence of approach to
the asymptotics. Note that such a scenario is an intrinsic feature
of the critical slowing down in random-anisotropy magnets. When
disorder is implemented by dilution of the non-magnetic component,
an approach of $z_{\rm eff}$ to its asymptotic value is not
necessarily only from above  \cite{Blavatska05}.

\section{Conclusions}\label{IV}

In this paper, we have analyzed the critical slowing down in
magnets influenced by random anisotropy. These magnets have a
second order phase transition to ferromagnetic order  for an
anisotropic ({\em cubic}) random axis distribution. Therefore, our goal
was to study relaxational dynamics of the non-conserved order
parameter in the vicinity of the phase transition point. For this
purpose we completed previous static RG 
calculations  \cite{Dudka05,RAM,Calabrese04} by calculating the two-loop
dynamical RG function $\zeta_{\Gamma}$ given in Eq. (\ref{main}).
Combining this result with the former data for the static critical
behaviour we obtained numerical values for the effective critical
exponent $z_{\rm eff}$ which governs the critical slowing down of
the relaxational time when $T_c$ is approached. In Figs.
\ref{zeff2},\ref{zeff} we give results for two most physically
interesting cases $m=2$ and $m=3$, which correspond to the
easy-plane and Heisenberg random anisotropy magnets.

Although the asymptotic dynamical critical behaviour of random
anisotropy systems with {\em cubic distribution}  is the same as
for the random-site Ising systems, the crossover between different
fixed points  considerably  influences the non-asymptotic critical
properties.  Different scenarios of dynamical critical behaviour
are observed. Since the approach to asymptotics is very slow, it
might be observed in real and numerical experiments. The effective
exponents measured may take values essentially differing form the
asymptotic  one (in our calculation $z=2.139$). For example in a
large region $z_{\rm eff}$ can be equal to the exponent of the
pure Ising model ($z=2.052$). One more particular feature of
critical slowing down in random anisotropy magnets which is
predicted by our analysis is that, contrary to the diluted
magnets, $z_{\rm eff}$ seems to reach its asymptotic value $z$ always
from the region $z_{\rm eff}>z$.

Another important contribution to the effective dynamical exponent could
come from the coupling of the order parameter to a conserved density
(changing from model A dynamics to model C  \cite{Halperin77}). Since
the stable fixed point has a non-diverging specific heat
the asymptotic discussed here would not be changed  \cite{dudka05}.
A more detailed account on that is in preparation.

It is our pleasure and honour to contribute this paper to the
conference on the occasion of Prof. I.R. Yukhnovskii 80th
birthday. His early work on the phase transitions theory and on an
account of the non-asymptotic criticality  \cite{Yukhnovskii}
preceded in many respects many later contributions to this field.
One of us (Yu.H.) is deeply indebted to the jubilee for
introducing into the fascinating field of phase transitions and
critical phenomena. R. F. acknowledges  the fruitful cooperation
with the Institute for Condensed Matter Physics.

This work was supported by Austrian Fonds zur F\"orderung der
wi\-ssen\-schaft\-li\-chen Forschung under Project No. P16574.

%
%% If you have problems with typesetting in ukrainian uncomment lines below.
%
  \label{last@page}
  \end{document}